\documentclass[]{pasj01}
\begin{document} 
\Received{2018/05/01}%{yyyy/mm/dd}
\Accepted{2018/06/01}%{yyyy/mm/dd}
%\Published{yyyy/mm/dd}

\title{
VERA monitoring of the radio jet 3C 84 during 2007--2013:
detection of non-linear motion}

%%% begin:list of authors
% Do NOT capitalize all letters in "textsc".
%\author{A-Firstname \textsc{A-Familyname}\altaffilmark{1}%
%\thanks{Example: Present Address is xxxxxxxxxx}}
%\altaffiltext{1}{A-Address of Institute}
%\email{aaaaa@xxx.xxx.xx.xx}

%\author{B-Firstname \textsc{B-Familyname},\altaffilmark{2}}
%\altaffiltext{2}{B-Address of Institute}
%\email{bbbbb@xxx.xxx.xx.xx}

%\author{C-Firstname \textsc{C-Familyname}\altaffilmark{3}}
%\altaffiltext{3}{C-Address of Institute}
%\email{ccccc@xxx.xxx.xx.xx}

\author{%
Koichiro \textsc{Hiura},\altaffilmark{1}
Hiroshi \textsc{Nagai},\altaffilmark{2}
Motoki \textsc{Kino},\altaffilmark{2,3}
Kotaro \textsc{Niinuma},\altaffilmark{4}
Kazuo \textsc{Sorai},\altaffilmark{1,5,6,7}
Hikaru \textsc{Chida},\altaffilmark{8}
Kazunori \textsc{Akiyama},\altaffilmark{9,10, 2}
Filippo \textsc{D'Ammando},\altaffilmark{11,12}
Gabriele \textsc{Giovannini},\altaffilmark{11,12}
Marcello \textsc{Giroletti},\altaffilmark{11}
Kazuhiro \textsc{Hada},\altaffilmark{2,13}
Mareki \textsc{Honma},\altaffilmark{2,13}
Shoko \textsc{Koyama},\altaffilmark{14}
Monica \textsc{Orienti},\altaffilmark{11}
Gabor \textsc{Orosz},\altaffilmark{15,16}
and
Satoko \textsc{Sawada-Satoh}\altaffilmark{15}}

\altaffiltext{1}{Department of Cosmosciences, Graduate School of Science, Hokkaido University, Kita 10, Nishi 8, Kita-ku, Sapporo, Hokkaido 060-0810, Japan}
\altaffiltext{2}{National Astronomical Observatory of Japan, 2-21-1 Osawa, Mitaka, Tokyo 181-8588, Japan}
\altaffiltext{3}{Kogakuin University, Academic Support Center, 2665-1 Nakano, Hachioji, Tokyo 192-0015, Japan}
\altaffiltext{4}{Graduate School of Sciences and Technology for Innovation, Yamaguchi University, 1677-1 Yoshida, Yamaguchi, Yamaguchi 753-8512, Japan}
\altaffiltext{5}{Department of Physics, Faculty of Science, Hokkaido University, Kita 10, Nishi 8, Kita-ku, Sapporo, Hokkaido 060-0810, Japan}
\altaffiltext{6}{Division of Physics, Faculty of Pure and Applied Sciences, University of Tsukuba, 1-1-1 Tennodai, Tsukuba, Ibaraki 305-8571, Japan}
\altaffiltext{7}{Tomonaga Center for the History of the Universe (TCHoU), University of Tsukuba, Tsukuba, Ibaraki 305-8571, Japan}
\email{sorai@astro1.sci.hokudai.ac.jp}
\altaffiltext{8}{Department of Physics, Tokai University, 4-1-1 Kitakaname, Hiratsuka-shi, Kanagawa 259-1292, Japan}
\altaffiltext{9}{National Radio Astronomy Observatory, 520 Edgemont Rd, Charlottesville, VA, 22903, USA}
\altaffiltext{10}{Massachusetts Institute of Technology, Haystack Observatory, Westford, MA 01886, USA}
\altaffiltext{11}{INAF Istituto di Radioastronomia, via Gobetti 101, I-40129 Bologna, Italy}
\altaffiltext{12}{Dipartimento di Fisica e Astronomia, Universit\`{a} di Bologna, Via Gobetti, 93/2, I-40127 Bologna, Italy}
\altaffiltext{13}{Department of Astronomical Science, The Graduate University for Advanced Studies (SOKENDAI), 2-21-1 Osawa, Mitaka, Tokyo 181-8588, Japan}
\altaffiltext{14}{Institute of Astronomy and Astrophysics, Academia Sinica, P.O. Box 23-141, Taipei 10617, Taiwan}
\altaffiltext{15}{Graduate School of Science and Engineering, Kagoshima University, 1-21-35 Korimoto, Kagoshima 890-0065, Japan}
\altaffiltext{16}{Xinjiang Astronomical Observatory, Chinese Academy of Sciences, 150 Science 1-Street, Urumqi, Xinjiang 830011, China}
%%% end:list of authors

%% `\KeyWords{}' always has to be placed before `\maketitle'.
 %Do NOT move this preamble from here!

\maketitle

\begin{abstract}
We present a kinematic study of the subparsec-scale radio jet of the radio galaxy 3C 84/NGC 1275 with the VLBI Exploration of Radio Astrometry (VERA) array at 22 GHz for 80 epochs from 2007 October to 2013 December. 
The averaged radial velocity of the bright component ``C3'' with reference to the radio core 
is found to be $0.27\pm0.02c$ between 2007 October and 2013 December.
This constant velocity of C3 is naturally 
explained by the advancing motion of the head of the mini-radio lobe.
We also find a non-linear component in the motion of C3 with respect to the radio core.
We briefly discuss possible origins of this non-linear motion.

\end{abstract}

\section{Introduction}
Radio-loud active galactic nuclei (AGNs) often have relativistic jets emanating from the vicinities of their central supermassive black holes (SMBHs).
Radio galaxies are thought to be misaligned radio-loud AGNs within the unified model of AGN \citep{Urry+Padovani1995}.
Thus, radio galaxies are ideal sources to explore the general properties of AGN jets since the misalignment of the jet axis with the line of sight provides a detailed view of the structure in the jet.

The bright radio source 3C 84 is associated with the giant elliptical galaxy NGC 1275 ($z = 0.0176$; \cite{Petrosian+2007}), which is a dominant member of the Perseus cluster.
Its proximity allows us to investigate not only its large-scale structures, but also the central subparsec-scale region, where the jet nozzle is located, with the high angular resolution provided by observations with Very Long Baseline Interferometry (VLBI).
Therefore, 3C 84 is an ideal source to study the formation mechanism of relativistic jets powered by an SMBH and the interaction between the jets and ambient medium in the vicinity of the SMBH (e.g., \cite{Giovannini+2018} and references therein).

3C 84 is an uncommon source exhibiting intermittent jet activity.
Its radio morphology has multiple lobe-like features with different position angles on broad spatial scales from pc to $\sim$ 10 kpc (e.g., \cite{Pedlar+1990,Walker+2000}).
3C 84 also shows pairs of 100 kpc-scale X-ray bubbles misaligned with each other \citep{Dunn+2006}.

In the central 5--10\ pc scale region, 3C 84 has two-sided compact radio jets/lobes, which were probably formed by the jet activity originating in the 1959 flare (\cite{Vermeulen+1994}; \cite{Walker+1994},\ \yearcite{Walker+2000}; \cite{Asada+2006}).
The morphology of 3C 84 is similar to Compact Symmetric Objects (CSOs: \cite{Readhead+1996}) as well as Fanaroff-Riley type-I radio galaxies (e.g., \cite{Dhawan+1998}).
Despite the CSO-like morphology, it is not a genuine young radio source because of the presence of large scale morphology.
Using the Very Long Baseline Array (VLBA) observations at 43\ GHz in the 1990's, \citet{Dhawan+1998} revealed that the inner 0.5 pc of the core has bright knots located along a line with multiple sharp bends.
These bends may reflect a precessing jet nozzle, or three-dimensional hydrodynamic Kelvin-Helmholtz instabilities in 3C 84 \citep{Dhawan+1998}, but no one has directly observed the wobbling motion of any particular component.

3C 84 did not undergo significant enhancement in the jet activities in the central sub-parsec region between 1959 and the early 2000s, suggested by observations showing a monotonic decrease in its radio flux density.
However, monitoring observations at 14.5\ GHz with a single-dish radio telescope at the University of Michigan Radio Astronomy Observatory (UMRAO) have detected brightening, starting from 2005 \citep{Abdo+2009}. 
In fact, the Monitoring Of Jets in Active galactic nuclei with
VLBA Experiments (MOJAVE; Lister et al. 2009) 15 GHz
VLBA observations of 3C 84, taken simultaneously with the Fermi Gamma-ray Space Telescope 
on 2008 August 25, show a significant brightening of the central
sub-parsec-scale structure, indicating that a flare is happening
in the innermost jet region. This brightening might be connected to the gamma-ray activity \citep{Abdo+2009}. 
Using the VLBI Exploration of Radio Astrometry (VERA), \citet{Nagai+2010} found that the brightening was ascribed to the central subparsec-scale core, accompanying the ejection of a new bright radio component (C3).
Therefore, 3C 84 is an adequate source for studying ongoing recurrent jet activity in the central subparsec-scale core. 
Using  observations at a higher spatial resolution with the VLBA at 43\ GHz, \citet{Suzuki+2012} found that C3 emerged from the radio core (C1) before 2005, and traveled southward following a parabolic trajectory on the celestial sphere.
\citet{Suzuki+2012} also found that the apparent speed of C3 with reference to C1 shows moderate sub-relativistic acceleration from $0.10c$ to $0.47c$ between 2003 November and 2008 November.

In order to understand the formation mechanism of jets, it is important to study the kinematic properties in the vicinity of the jet's base.
In this paper, we present the detailed kinematics of C3 to reveal its true nature.
We investigate the kinematics of C3 in detail by monitoring the subsequent motion of the non-linear trajectory found by \citet{Suzuki+2012}.
In order to confirm the nature of C3, we will also discuss it by approaching from light curve in a forthcoming paper.
Note taht the redshift of 3C 84 corresponds to an angular scale of $0.353\ \mathrm{pc\ mas^{-1}}$ ($0.1\ \mathrm{mas\ yr^{-1}} = 0.115c$) assuming $H_{0}=71\ \mathrm{km\ s^{-1}\ Mpc^{-1}}$, $\Omega_{\mathrm{M}}=0.27$, and $\Omega_{\mathrm{\Lambda}}=0.73$ \citep{Komatsu+2009}.

\section{Observation and Data Reduction}
\subsection{VERA data at 22\ GHz}
In order to investigate the detailed kinematics of C3, we mainly used the GENJI programme (Gamma-ray Emitting Notable AGN Monitoring with Japanese VLBI; \cite{Nagai+2013}) data at 22\ GHz (2010 November -- 2013 December, 68 epochs).
The GENJI programme aims for dense sampling of $\gamma$-ray loud AGNs using the available calibrator time in the Galactic maser astrometry project of VERA.
Maser sessions need to monitor a bright calibrator once in every $\sim80$\ minutes, for which we use GENJI sources including 3C 84.
One of the goals of the GENJI programme is to identify the radio counterparts of gamma-ray emitting regions in AGN, by comparing radio and gamma-ray light curves.
We also aim to study the kinematics of the jet.
We pay attention to the time variations in the flux density on a time scale shorter than one month, which provides quick follow-up observations after $\gamma$-ray flares.   
Thanks to this dense monitoring, we can obtain detailed data of 3C 84 on subparsec scales.

VERA consists of four stations with a maximum baseline length of $\sim2,270$\ km.
This corresponds to a typical angular spatial resolution of $\sim1$\ mas. 
In addition to the GENJI programme data, we also used published data (\cite{Nagai+2010}, 2007 October - 2008 May, 7 epochs) and archival VERA data (2009 February - 2010 February, 5 epochs) at 22\ GHz.
During each observation, total on-source time for 3C 84 was typically 30 minutes, consisting of 4-6 scans at different hour angles.
Data reduction was performed using the National Radio Astronomy Observatory (NRAO) Astronomical Imaging Processing System (AIPS) in the same way as \citet{Nagai+2013}.
The final images were obtained after a number of iterations with \textit{modelfit} and self-calibration implemented in the Difmap software package \citep{Shepherd+1994}. 
Our final dataset includes data from 80 epochs at a subparsec scale (table\ \ref{tab:obs_information}, see supplementary table\  \ref{tab:obs_information}).

% supplementary table: Epoch, rms, synthesized beam size, peak brightness, and total model-fitted flux for all images. \label{tab:obs_information} (supplementary_table1.tex)
% Please comment out the line "\input{supplementary_table1}" (line 519) except for the electric edition.

\begin{longtable}{lrrlrr}
\caption{Epoch, rms, synthesized beam size, peak brightness, and total model-fitted flux for all images. See supplementary table\ \ref{tab:obs_information}.}
\label{tab:obs_information}
\hline\hline
\multicolumn{2}{c}{Epoch} & \multicolumn{1}{c}{Image noise rms} & \multicolumn{1}{c}{Beam\footnotemark[$\dagger$]} & \multicolumn{1}{c}{$I_{\mathrm{peak}}$\footnotemark[$\ddagger$]} & \multicolumn{1}{c}{$S_{\mathrm{total}}$\footnotemark[$\S$]} \\
\multicolumn{1}{c}{Date} & MJD-54397\footnotemark[$*$] & ($\mathrm{mJy\ beam^{-1}}$) & (mas $\times$ mas, deg) & ($\mathrm{Jy\ beam^{-1}}$) & \multicolumn{1}{c}{(Jy)} \\
\hline
\endhead
\hline
\endfoot
\hline
\multicolumn{6}{@{}l@{}}{\hbox to 0pt{\parbox{140mm}{\footnotesize
Notes.
\par\noindent
\footnotemark[$*$] Time gap between Modified Julian Date (MJD) of the epoch and MJD 54397 (2007 Oct.\,24).
\par\noindent
\footnotemark[$\dagger$] Major axis, minor axis, and position angle of synthesized beam.
\par\noindent
\footnotemark[$\ddagger$] Peak brightness for each image.
\par\noindent
\footnotemark[$\S$] Total model-fitted flux and its error for each image.
The amplitude calibration error is assumed to be 10\% of flux density, according to a number of experiences using VERA (e.g., \cite{Petrov+2012}).
}\hss}}
\endlastfoot
2007 Oct.\,24 & 0 & 27.9 & $1.29\times0.76,\,-54.7$ & 3.5 & $8.4\pm0.8$ \\
2007 Nov.\,20 & 27 & 42.0 & $1.17\times0.83,\,-53.4$ & 4.4 & $9.9\pm1.0$ \\
2007 Dec.\,27 & 64 & 27.1 & $1.24\times0.73,\,-52.8$ & 4.1 & $10.5\pm1.0$ \\
2008 Feb.\,4 & 103 & 34.9 & $1.27\times0.79,\,-58.1$ & 4.1 & $9.9\pm1.0$ \\
2008 Mar.\,3 & 131 & 38.0 & $1.18\times0.75,\,-49.5$ & 3.7 & $9.0\pm0.9$ \\
\end{longtable}

\subsection{Gaussian Model Fitting}
\label{sec:gaussian_model_fitting}
In order to quantify the position, size, flux density of bright regions in 3C 84 on subparsec scales, 
we performed the standard  model fitting procedure by employing the task \textit{modelfit} in Difmap. 
Unlike \citet{Nagai+2010}, we adopted only circular Gaussian model components (not elliptical one), to the visibility data of each epoch in order to avoid extremely elongated components and to facilitate comparison of the features and their identification (e.g., \cite{Kudryavtseva+2011}). 
We judged the goodness of the fit from reduced $\chi^{2}$ statistics.

Resultant images for the epochs before 2010 December were well represented by 3 major components (C1, C2, and C3; shown in section\ \ref{sec:total_intensity_image}), which were the same components as identified in \citet{Nagai+2013}.
In addition to these three components, there was additional emission bridging C1 and C3, for the epochs after 2010 December.
This additional emission was also detected in 43\ GHz VERA observations \citep{Nagai+2012}.
Then, we modeled this bridging emission by using a circular Gaussian component (C4) for the epochs after 2010 December.
The choice of 3 or 4 components was verified by \textit{F}-test across all epochs. 
  
\subsection{Positional Accuracy}
\label{positional_accuracy}
\begin{figure}
\begin{center}
\includegraphics[width=80mm]{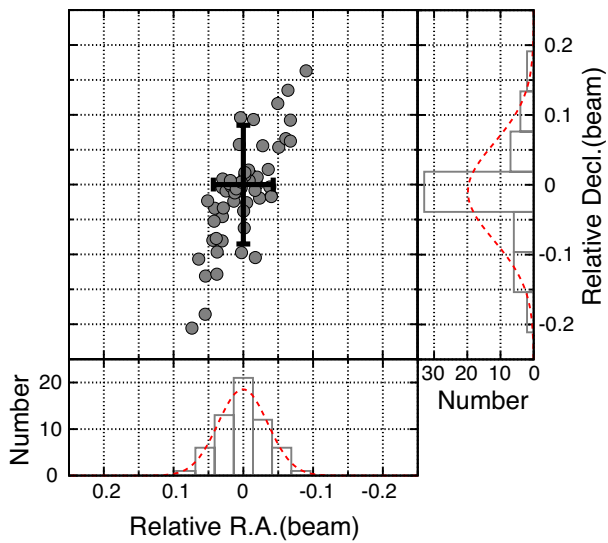}
\end{center}
\caption{Estimation of positional error for C3.
Each point represents the difference of C3 position with reference to 
C1 measured between two epochs close in time (within 30 days).
These are normalized with the beam size, which are averaged of the FWHM on the major-minor axes
for the synthesized beam over two adjacent epochs. 
Thick bars correspond to the positional error for right ascension and declination, and are estimated to be $0.043\ \theta_{\mathrm{beam}}$ and $0.085\ \theta_{\mathrm{beam}}$, respectively. 
Histograms of the difference of C3 position are shown along each axis.
Red dashed lines represents normal distribution functions with the means
and unbiased standard deviations of samples along each axis.
Areas of under the normal distribution functions correspond to those of the histograms.}
\label{fig:positional_accuracy}
\end{figure}

It is important to check the positional accuracy for studying the detailed kinematics.
In the same manner as \citet{Suzuki+2012}, we estimated the C3 positional errors by examining the scatter in the C3 positions with reference to the optically thick component C1 in images between two close epochs (within 30 days separation), such that the source structures are approximately same in both epochs. 
Assuming that the apparent motion of C3 is $0.5c$, this motion corresponds to a 0.036\ mas positional change ($<0.05$ of typical VERA 22\ GHz beam) within 30 days. 
We have analyzed 60 pairs of images and figure\ \ref{fig:positional_accuracy} shows the differences of relative positions of C3 with respect to C1 for these 60 pairs (table\ \ref{tab:positional_accuracy}, see supplementary table\ \ref{tab:positional_accuracy}).
Each data point in figure\ \ref{fig:positional_accuracy} is normalized with the beam size ($\theta_{B}^{\mathrm{mean}}$), which is averaged of the FWHM on the major-minor axes for the synthesized beam over two adjacent epochs.
The unbiased standard deviations along right ascension ($\sigma_{s}^{\mathrm{R.A.}}$) and declination ($\sigma_{s}^{\mathrm{Decl.}}$) normalized by $\theta_{B}^{\mathrm{mean}}$ are 0.035 and 0.069, respectively. 
As shown in figure 1, we can conservatively regard histograms of the difference of C3 positions as normally-distributed, 
since most bins of the histograms are covered by the normal distribution functions, 
especially in the tails of the distributions.
Thus, assuming that each point in figure\ \ref{fig:positional_accuracy} is normally-distributed, the $100\,(1-\alpha)$\% confidence interval of the standard deviation of population for statistical ensemble \textit{i}, $\sigma_{p}^{i}$, is estimated from \textit{N} samples as
\begin{equation}
\frac{(N-1){\sigma_{s}^{i}}^{2}}{\chi^{2}_{N-1}(\alpha/2)} \leq {\sigma_{p}^{i}}^{2} \leq \frac{(N-1){\sigma_{s}^{i}}^{2}}{\chi^{2}_{N-1}(1-\alpha/2)},
\end{equation}
where $\chi^{2}_{N-1}(\alpha)$ is the $\chi^{2}$ statistic for degrees of freedom (dof) $=N-1$ on which the event $\chi^{2} \geq \chi^{2}_{N-1}$ happens with probability $\alpha$.
We apply this estimator for the right ascension ensemble ($i=\mathrm{R.A.}$) and the declination ensemble ($i=\mathrm{Decl.}$).
Given $N=60$ and $\alpha=0.05$, the standard deviations of the populations for both right ascension ($\sigma_{p}^{\mathrm{R.A.}}$) and declination ($\sigma_{p}^{\mathrm{Decl.}}$) are estimated to be
\begin{equation}
0.030 \leq \sigma_{p}^{\mathrm{R.A.}} \leq 0.043,
\end{equation}
\begin{equation}
0.059 \leq \sigma_{p}^{\mathrm{Decl.}} \leq 0.085.
\end{equation}
Hereafter, the positional accuracy of C3 is conservatively set as $0.043\ \theta_{\mathrm{beam}}$ for right ascension and $0.085\ \theta_{\mathrm{beam}}$ for declination, where $\theta_{\mathrm{beam}}$ is the beam size, which is  averaged of the FWHM on the major-minor axes for the synthesized beam in each epoch.
In the same way as mentioned above, the positional errors of C2 and C4 are also estimated.
We set the positional accuracy of C2 as $0.14\ \theta_{\mathrm{beam}}$ for right ascension and $0.13\ \theta_{\mathrm{beam}}$ for declination, and that of C4 as $0.077\ \theta_{\mathrm{beam}}$ for right ascension and $0.082\ \theta_{\mathrm{beam}}$ for declination.

% supplementary table: Dispersions of C3 position with reference to C1. \label{tab:positional_accuracy} (supplementary_table2.tex)
% Please comment out the line "\input{supplementary_table2}" (line 520) except for the electric edition.

\begin{longtable}{lcrr}
\caption{Dispersions of C3 position with reference to C1. See supplementary table\ \ref{tab:positional_accuracy}.}
\label{tab:positional_accuracy}
\hline\hline
\multicolumn{1}{l}{Pairs\footnotemark[$*$]} & $\theta_{B}^{\mathrm{mean}}$\footnotemark[$\dagger$] & $\mathit{\Delta}\mathrm{R.A.}$\footnotemark[$\ddagger$] & $\mathit{\Delta}\mathrm{Decl.}$\footnotemark[$\ddagger$] \\
 & (mas) & (beam) & (beam) \\
\hline
\endhead
\hline
\endfoot
\hline
\multicolumn{4}{@{}l@{}}{\hbox to 0pt{\parbox{85mm}{\footnotesize
Notes.
\par\noindent
\footnotemark[$*$] Pairs of adjacent epochs with their separation $\leq 30$\ days (for which the motion of components are negligible).
\par\noindent
\footnotemark[$\dagger$] Beam sizes averaged for the major-minor axis and two epochs.
\par\noindent
\footnotemark[$\ddagger$] Differences of relative positions in right ascension and declination of C3 with reference to C1 between two epochs. These are normalized by $\theta_{B}^{\mathrm{mean}}$.%the beam sizes averaged between the pairs.
}\hss}}
\endlastfoot
2007 Oct.\,24/2007 Nov.\,20 & 1.01 & $-0.005$ & $0.022$ \\
2008 Feb.\,4/2008 Mar.\,3 & 0.99 & $-0.081$ & $-0.008$ \\
2010 Nov.\,28/2010 Nov.\,29 & 1.04 & $0.100$ & $-0.032$ \\
2010 Nov.\,29/2010 Dec.\,4 & 1.05 & $-0.006$ & $0.015$ \\
2011 Jan.\,11/2011 Jan.\,29 & 0.99 & $0.098$ & $0.017$ \\
\end{longtable}

%%%%%%%%%%%%%%%%

\section{Results}
\subsection{Total Intensity Image} 
\label{sec:total_intensity_image}
\begin{figure}
\begin{center}
\includegraphics[width=80mm]{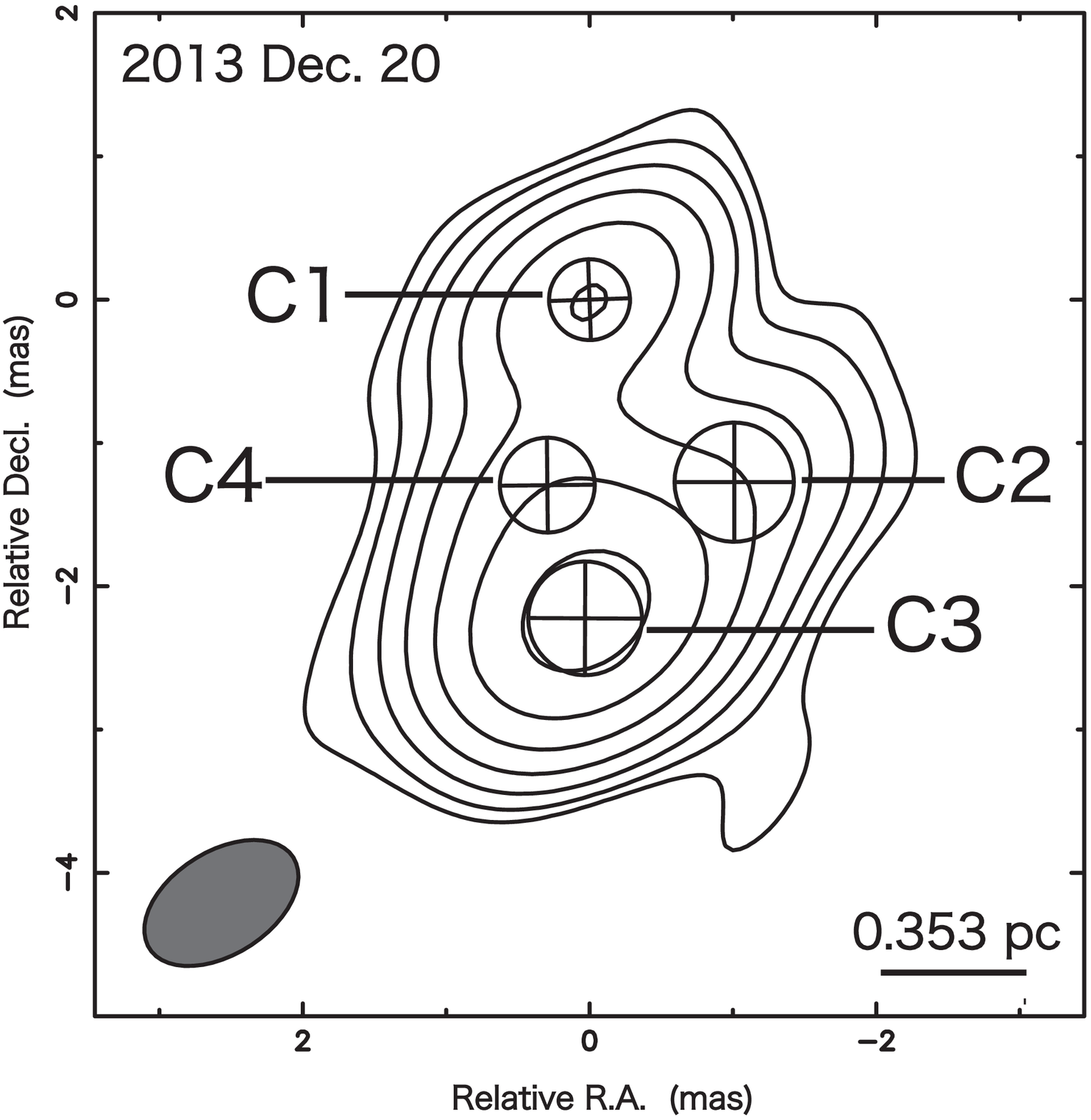}
\end{center}
\caption{22\ GHz total intensity image of 3C 84 on 2013 December 20 (MJD 56646) with circular Gaussian components imposed.
The diameters of these circular components represent the full width at the half maximum (FWHM) sizes of the individual fitted Gaussian components. 
The ellipse shown at the bottom left corner of the image indicates the FWHM of the convolved beam.
The FWHM of the convolved beam is $1.19 \times 0.72$\ mas at the position angle of \timeform{-57.6D}
The contours are plotted at the level of $3\sigma \times 2^{n}\ (n=0,\,1,\,2,\,3, \cdots)$, where $\sigma$ is image noise rms of $40.6\ \mathrm{mJy\ beam^{-1}}$.
The thick bar at the bottom right corner corresponds to 0.353 pc.}
\label{fig:total_intensity_image}
\end{figure}

As an example, Figure\ \ref{fig:total_intensity_image} shows the self-calibrated image of 3C 84 on 2013 December 20.
As described in section\ \ref{sec:gaussian_model_fitting}, the subparsec-scale structure in 3C 84 can be represented by 3 (4) circular Gaussian components for the epochs before (after) 2010 December.
Due to a lack of short baselines, we only detected the structure within $\sim 3$\ mas from the phase tracking center, but missed extended structures.
The jet extends southward from the northern bright core component C1.
C2 is $\sim1.6$\ mas away from C1, and its position angle relative to C1 (from north to east) is $\sim\timeform{218D}$ on 2013 December 20.
C3 and C4 are located at $\sim2.2$\ mas, $\sim\timeform{179D}$\ and $\sim1.3$\ mas, $\sim\timeform{167D}$ relative to C1, respectively.
No counter jet component is detected at a level of $3\sigma$ throughout all epochs. 
Physical parameters of all fitted components over 80 epochs are listed in table \ \ref{tab:relative_position}, \ref{tab:size}, and \ref{tab:flux} (see supplementary table\ \ref{tab:relative_position}, \ref{tab:size}, and \ref{tab:flux}).

% supplementary table: Relative positions from C1. \label{tab:relative_position} (supplementary_table3.tex)
% supplementary table: FWHM size of fitted components. \label{tab:size} (supplementary_table4.tex)
% supplementary table: Flux of fitted components. \label{tab:flux} (supplementary_table5.tex)
% Please comment out the line "\input{supplementary_table3}" (line 521), "\input{supplementary_table4}" (line 522), "\input{table5}" (supplementary_line 523) except for the electric edition.

\begin{longtable}{lrrrrrr}
\caption{Relative positions from C1. See supplementary table\ \ref{tab:relative_position}.}
\label{tab:relative_position}
\hline\hline
\multicolumn{1}{l}{Epoch} & \multicolumn{2}{c}{C2} & \multicolumn{2}{c}{C3} & \multicolumn{2}{c}{C4} \\
 & \multicolumn{1}{c}{$\mathit{\Delta}\mathrm{R.A.}$\footnotemark[$*$]} & \multicolumn{1}{c}{$\mathit{\Delta}\mathrm{Decl.}$\footnotemark[$*$]} & \multicolumn{1}{c}{$\mathit{\Delta}\mathrm{R.A.}$\footnotemark[$\dagger$]} & \multicolumn{1}{c}{$\mathit{\Delta}\mathrm{Decl.}$\footnotemark[$\dagger$]} & \multicolumn{1}{c}{$\mathit{\Delta}\mathrm{R.A.}$\footnotemark[$\ddagger$]} & \multicolumn{1}{c}{$\mathit{\Delta}\mathrm{Decl.}$\footnotemark[$\ddagger$]} \\
 & \multicolumn{1}{c}{(mas)} & \multicolumn{1}{c}{(mas)} & \multicolumn{1}{c}{(mas)} & \multicolumn{1}{c}{(mas)} & \multicolumn{1}{c}{(mas)} & \multicolumn{1}{c}{(mas)} \\
\hline
\endhead
\hline
\endfoot
\hline
\multicolumn{7}{@{}l@{}}{\hbox to 0pt{\parbox{160mm}{\footnotesize
Notes.
\par\noindent
Positional error is estimated in section\ \ref{positional_accuracy}.
\par\noindent
\footnotemark[$*$] Relative right ascension and declination between C1 and C2.
\par\noindent
\footnotemark[$\dagger$] Relative right ascension and declination between C1 and C3.
\par\noindent
\footnotemark[$\ddagger$] Relative right ascension and declination between C1 and C4.
}\hss}}
\endlastfoot
2007 Oct.\,24 & $-0.701\pm0.145$ & $-1.276\pm0.136$ & $0.002\pm0.044$ & $-0.827\pm0.087$ & \multicolumn{1}{c}{$\cdots$} & \multicolumn{1}{c}{$\cdots$} \\
2007 Nov.\,20 & $-0.706\pm0.141$ & $-1.253\pm0.132$ & $0.033\pm0.043$ & $-0.874\pm0.085$ & \multicolumn{1}{c}{$\cdots$} & \multicolumn{1}{c}{$\cdots$} \\
2007 Dec.\,27 & $-0.646\pm0.138$ & $-1.246\pm0.130$ & $0.042\pm0.042$ & $-0.879\pm0.083$ & \multicolumn{1}{c}{$\cdots$} & \multicolumn{1}{c}{$\cdots$} \\
2008 Feb.\,4 & $-0.584\pm0.145$ & $-1.294\pm0.136$ & $0.072\pm0.044$ & $-0.886\pm0.087$ & \multicolumn{1}{c}{$\cdots$} & \multicolumn{1}{c}{$\cdots$} \\
2008 Mar.\,3 & $-0.664\pm0.136$ & $-1.302\pm0.127$ & $0.032\pm0.042$ & $-0.902\pm0.082$ & \multicolumn{1}{c}{$\cdots$} & \multicolumn{1}{c}{$\cdots$} \\
\end{longtable}

\begin{longtable}{lcccc}
\caption{FWHM size of fitted components. See supplementary table\ \ref{tab:size}.}
\label{tab:size}
\hline\hline
\multicolumn{1}{l}{Epoch} & \multicolumn{1}{c}{C1 (mas)} & \multicolumn{1}{c}{C2 (mas)} & \multicolumn{1}{c}{C3 (mas)} & \multicolumn{1}{c}{C4 (mas)} \\
\hline
\endhead
\hline
\endfoot
\hline
\multicolumn{5}{@{}l@{}}{\hbox to 0pt{\parbox{105mm}{\footnotesize
Notes.
\par\noindent
Error is $1\sigma$ level, and estimated in the same way as the estimation of positional error (see section\ \ref{positional_accuracy}).
The upper limit is $1\sigma$ level.
}\hss}}
\endlastfoot
2007 Oct.\,24 & $<0.24$ & $0.57\pm0.52$ & $0.42\pm0.05$ & $\cdots$ \\
2007 Nov.\,20 & $<0.24$ & $<0.50$ & $0.48\pm0.05$ & $\cdots$ \\
2007 Dec.\,27 & $<0.23$ & $0.65\pm0.49$ & $0.48\pm0.05$ & $\cdots$ \\
2008 Feb.\,4 & $<0.24$ & $0.67\pm0.51$ & $0.49\pm0.05$ & $\cdots$ \\
2008 Mar.\,3 & $<0.23$ & $0.51\pm0.48$ & $0.47\pm0.05$ & $\cdots$ \\
\end{longtable}

\begin{longtable}{lrrrc}
\caption{Flux of fitted components. See supplementary table\ \ref{tab:flux}.}
\label{tab:flux}
\hline\hline
\multicolumn{1}{l}{Epoch} & \multicolumn{1}{c}{C1} & \multicolumn{1}{c}{C2} & \multicolumn{1}{c}{C3} & \multicolumn{1}{c}{C4} \\
 & \multicolumn{1}{c}{(Jy)} & \multicolumn{1}{c}{(Jy)} & \multicolumn{1}{c}{(Jy)} & \multicolumn{1}{c}{(Jy)} \\
\hline
\endhead
\hline
\endfoot
\hline
\multicolumn{5}{@{}l@{}}{\hbox to 0pt{\parbox{105mm}{\footnotesize
Notes.
\par\noindent
Flux and its error for each fitted component. Error is estimated as the root sum squares of calibration error (10\% of component) and image noise rms of each epoch.
}\hss}}
\endlastfoot
2007 Oct.\,24 & $2.66\pm0.27$ & $2.13\pm0.21$ & $3.60\pm0.36$ & $\cdots$ \\
2007 Nov.\,20 & $2.94\pm0.29$ & $2.14\pm0.21$ & $4.83\pm0.48$ & $\cdots$ \\
2007 Dec.\,27 & $3.11\pm0.31$ & $2.81\pm0.28$ & $4.53\pm0.45$ & $\cdots$ \\
2008 Feb.\,4 & $2.85\pm0.29$ & $2.75\pm0.28$ & $4.29\pm0.43$ & $\cdots$ \\
2008 Mar.\,3 & $2.70\pm0.27$ & $2.05\pm0.21$ & $4.28\pm0.43$ & $\cdots$ \\
\end{longtable}

%%%%%%%%%%%%%%%%%%

\subsection{Change in Relative Positions of Components}
\label{change_in_relative_positions_of_components}
We need to define the reference position for the following discussion on kinematics, since the information of absolute position in each image is lost at the fringe-fitting and self-calibration processes.
In the same way as \citet{Suzuki+2012}, we set the optically thick radio core C1 (e.g., Hodgson et al. 2018) as the reference position, and evaluate kinematics of other components relative to it.

\begin{figure}
\begin{center}
\includegraphics[width=80mm]{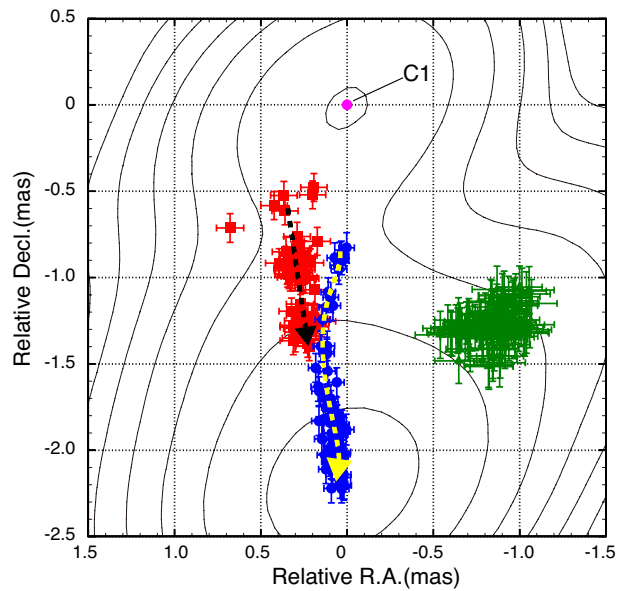}
\end{center}
\caption{Sky position plots of C2 (green triangles), C3 (blue circles), and C4 (red squares) for all 80 epochs, superposed on the contours of the 22\ GHz intensity distribution on 2013 December 20 (MJD 56646).
C1 (magenta circle) is set as the reference position at the origin.
Positional error is estimated in section\ \ref{positional_accuracy}.
Dashed curved and straight lines indicate the wobbling motion of C3 and the linear motion of C4, respectively.}
\label{fig:radec_map}
\end{figure}

\begin{figure}
\begin{center}
\includegraphics[width=85mm]{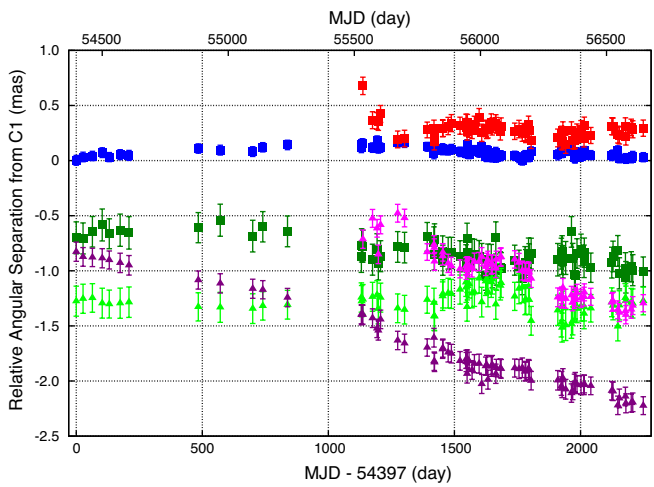}
\end{center}
\caption{Relative angular separation from C1 projected on right ascension (square symbols) and declination (triangle symbols) as a function of time in days over about 6 years.
Green squares and lime triangles represents relative separation between C1 and C2 projected on right ascension and declination, respectively.
Blue squares and purple triangles represents relative separation between C1 and C3 projected on right ascension and declination, respectively.
Red squares and magenta triangles represents relative separation between C1 and C4 projected on right ascension and declination, respectively.
Positional errors are estimated in section\ \ref{positional_accuracy}. 
Modified Julian Date (MJD) 54397 is 2007 October 24.}
\label{fig:distance_22G_all}
\end{figure}

Figure\ \ref{fig:radec_map} shows the evolution of the peak positions of C2, C3, and C4 with reference to C1 that were obtained from the total intensity images at 80 epochs over six years.
C2 existed before the advent of C3 \citep{Nagai+2010,Suzuki+2012}, and its motion is almost zero with no systematic changes.
On the other hand, C3 and C4 travel mainly southward with a small component in the east-west direction.
This trend in the motion of C3 is the same as reported in \citet{Suzuki+2012} for the period between 2003 November and 2008 November. 
The motions mentioned above of C2, C3 and C4 relative to C1 can be also seen in figure\ \ref{fig:distance_22G_all}.

\begin{figure}
\begin{center}
\includegraphics[width=90mm]{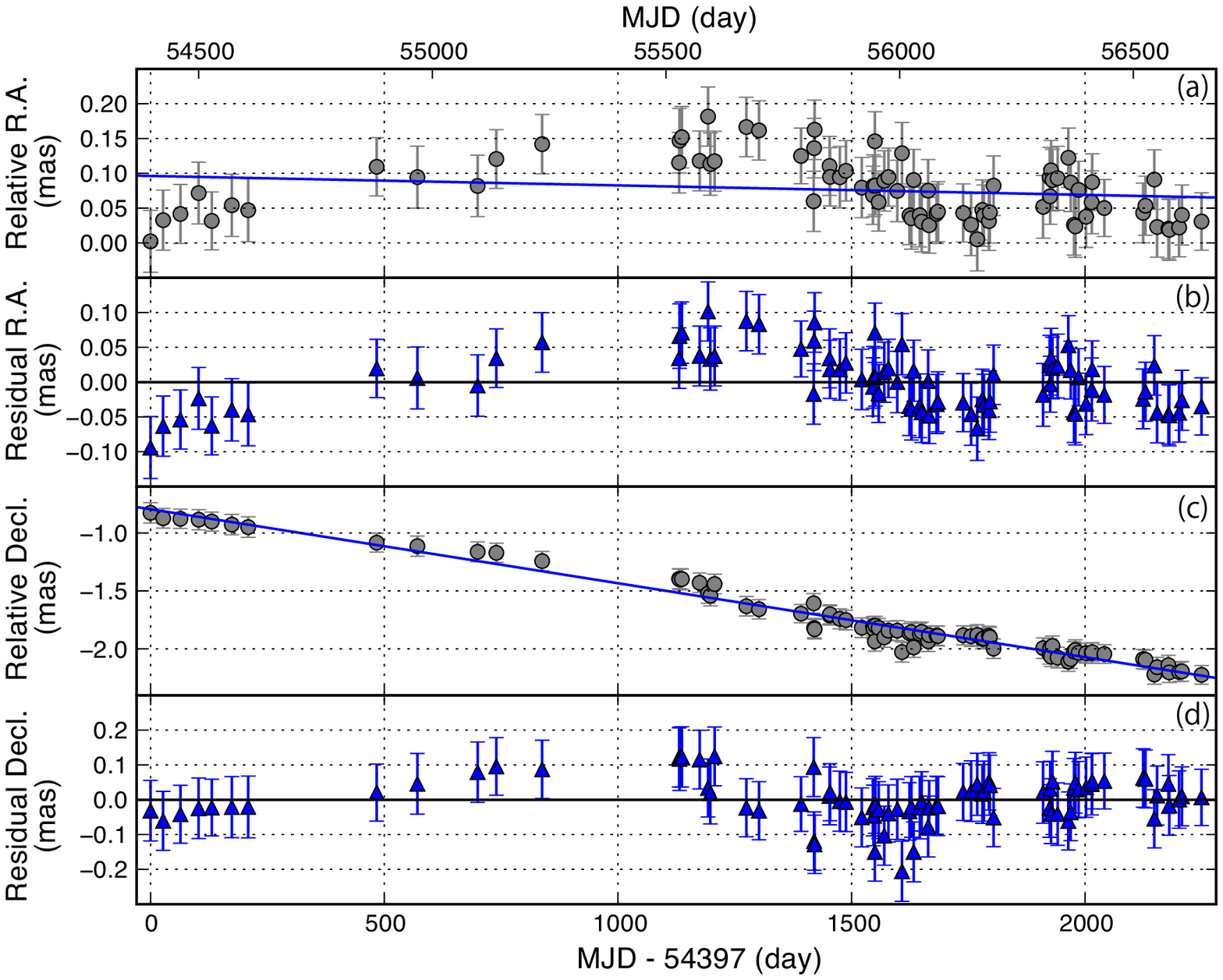}
\end{center}
\caption{(a) Relative angular separation projected on right ascension and (c) declination  between C1 and C3 as a function of time in days over about six years.
Blue lines are fitted linear functions.
Blue triangles in (b) and (d) are the residuals after subtracting the linear trends in (a) and (c), respectively.
Positional error is estimated in section\ \ref{positional_accuracy}. 
MJD 54397 is 2007 October 24.}
\label{fig:leastsq_C3_linear}
\end{figure}

In order to describe the average positional change of C3, we define the average proper motion as a vector $\left(\left\langle \mu \right\rangle,\ \left\langle \phi \right\rangle\right)$, where $\left\langle \mu \right\rangle$ represents the mean angular speed of motion and $\left\langle \phi \right\rangle$ expresses the average direction of motion relative to C1.
The average values are calculated as follows: $\left\langle \mu \right\rangle = \left(\left\langle \mu_{x} \right\rangle^{2} + \left\langle \mu_{y} \right\rangle^{2}\right)^{1/2}$ and $\left\langle \phi \right\rangle = \arctan \left(\left\langle \mu_{x} \right\rangle/\left\langle \mu_{y} \right\rangle \right)$, where $\left\langle \mu_{x} \right\rangle$ and $\left\langle \mu_{y} \right\rangle$ are the average angular speed projected on $x$ (Right Ascension) and $y$ (Declination) axes, respectively.
Then, we fit the $(x,\,y)$ position as a function of time for C3 with straight lines that minimize the $\chi^{2}$ statistic as presented in figure\ \ref{fig:leastsq_C3_linear}(a) and \ref{fig:leastsq_C3_linear}(c).
The best-fit values are $\left\langle \mu_{x} \right\rangle = -0.005 \pm 0.009\ \mathrm{mas\ yr^{-1}}$, $\left\langle \mu_{y} \right\rangle = -0.23 \pm 0.02\ \mathrm{mas\ yr^{-1}}$.
Therefore, the average proper motion vector is derived as $\left(\left\langle \mu \right\rangle,\ \left\langle \phi \right\rangle\right) = (0.23 \pm 0.02\ \mathrm{mas\ yr^{-1}},\,\timeform{181.2D} \pm \timeform{2.3D})$.
This average apparent speed corresponds to $\left\langle\beta_{\mathrm{app}}\right\rangle=0.27 \pm 0.02$ in units of the speed of light $c$, and is consistent with the result of \citet{Suzuki+2012} ($\left\langle\beta_{\mathrm{app}}\right\rangle=0.23 \pm 0.06$ between 2003 November 20 and 2007 November 2, when C3 is not identified at 22\ GHz images).
Assuming the jet viewing angle of \timeform{25D} adopted in \citet{Abdo+2009}, the intrinsic speed of C3 can be estimated as about $0.40c$, which corresponds to the Lorentz factor of about 1.1 (Doppler factor $\sim 1.4$).

We also derive the average proper motion vector of C4, and $\left(\left\langle \mu \right\rangle,\ \left\langle \phi \right\rangle\right) = (0.27 \pm 0.05\ \mathrm{mas\ yr^{-1}},\,\timeform{188.4D} \pm \timeform{9.5D})$ (see dashed straight line shown in figure\ \ref{fig:radec_map}).
This mean angular speed is equivalent to $\left\langle\beta_{\mathrm{app}}\right\rangle=0.31 \pm 0.05$, and the intrinsic speed, the Lorentz factor and the Doppler factor can be estimated $\sim0.44c$, $\sim1.1$ and $\sim1.5$, respectively, assuming the jet viewing angle of \timeform{25D}.
These values of C4 are similar to those of C3.

The uncertainties in the best-fit parameters can be estimated from the confidence interval, which is generally derived with grid-search techniques of the $\chi^2$ surface.
However, the periodical motion model examined in section\ \ref{sec:wobbling} has nine parameters, making a grid-search computationally expensive and challenging.
Instead, in this paper, we derived estimates of uncertainties in the best-fit parameters of the proper motion models with a Monte Carlo simulation as follows.
We created $10^5$ trial data sets generated from the best-fit model with Gaussian random noises, where the standard variations are same to the positional errors derived in section\ \ref{positional_accuracy}.
Samples of the best-fit parameters for all trial data sets were derived with the least-square method and were used to estimate confidence limits.
We took the edges of the middle 99.7\% (3$\sigma$) fraction of the samples, and adopted them as estimates of the $3\sigma$-confidence interval. We note that the derived uncertainties are larger than the standard errors of the least square fitting, indicating that the derived uncertainties are more robust than the standard errors.

\subsection{Periodicity Analysis on Wobbling Motion of C3}
\label{sec:wobbling}
After subtracting the linear trends from the C3 positional change in figure\ \ref{fig:leastsq_C3_linear}(a) and \ref{fig:leastsq_C3_linear}(c), the residual positional changes indicate oscillatory behavior as shown in figure\ \ref{fig:leastsq_C3_linear}(b) and \ref{fig:leastsq_C3_linear}(d).
This indication is verified as follows.

\begin{figure}
\begin{center}
\includegraphics[width=90mm]{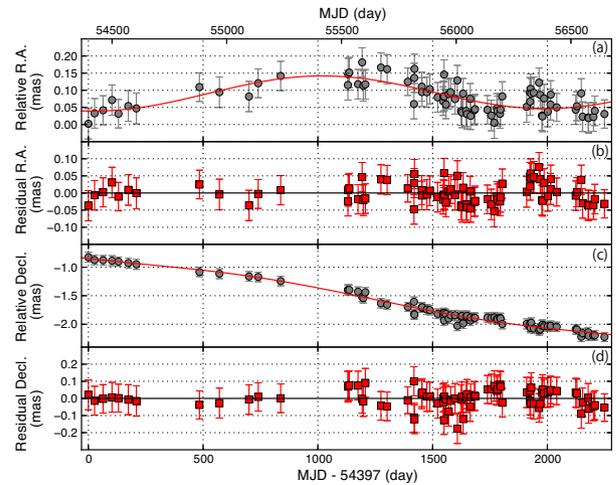}
\end{center}
\caption{(a) Relative angular separation projected on right ascension and (c) declination  between C1 and C3 as a function of time over about six years.
Red lines are fitted periodic functions with the common period between (a) and (c).
Red squares in (b) and (d) are the residuals after subtracting the periodic trends in (a) and (c), respectively.
Positional errors are estimated in section\ \ref{positional_accuracy}. 
MJD 54397 is 2007 October 24.}
\label{fig:leastsq_C3_periodic}
\end{figure}

We performed two types of analyses in order to check whether the relative motion of C3 with reference to C1 has periodicity.
First, we examined the significance of periodicity by fitting the $(x,\,y)$ positional change of C3 using two types of functions that minimize the $\chi^{2}$ statistic (figure\ \ref{fig:leastsq_C3_linear} and \ref{fig:leastsq_C3_periodic}) and employe some information criteria.
One is described by a linear motion model as
\begin{equation}
x(t) = at + b,
\end{equation}
\begin{equation}
y(t) = ct + d,
\end{equation}
and the other is expressed by a periodic motion model as
\begin{equation}
x(t) = A\sin \left(\frac{2\pi}{P_{\mathrm{obs}}^{\mathrm{fit}}}t + \frac{\pi}{180}B\right) + Ct + D,
\end{equation}
\begin{equation}
y(t) = E\sin \left(\frac{2\pi}{P_{\mathrm{obs}}^{\mathrm{fit}}}t + \frac{\pi}{180}F\right) + Gt + H,
\end{equation}
where the units of $x(t)$ and $y(t)$ are in mas, $t$ is the time from 2007 October 24 (MJD 54397) in days, and $P^{\mathrm{obs}}_{\mathrm{fit}}$ is the common parameter which denotes the period of the periodic function.

The best-fit parameters are listed in table\ \ref{tab:best_linear} and \ref{tab:best_periodic}.
The errors of these parameters were $3\sigma$-confidence intervals estimated by the Monte Carlo method with $10^{5}$ trials, considering correlation between parameters. 
Assuming that underlying errors of data points are normally distributed and independent, the reduced $\chi^{2}$ of the best fit based on the linear motion model and the periodic motion model are 0.76 and 0.45, respectively.  
Hence, we selected the more probable one of these two best-fit models using \textit{F}-test and the Akaike information criterion (AIC) \citep{Akaike1974}, the latter is an indicator of the relative quality of a statistical model for a given dataset.
The preferred model verified by these two methods is the periodic motion model.
Comparing the AIC values of these two models, we derived the relative probability of the linear motion model to the periodic motion model to be $4.2\times 10^{-9}$, which strongly suggests that the motion is periodic.

% table: The best-fit parameters of the linear motion model. \label{tab:best_linear} (table6.tex)
% table: The best-fit parameters of the periodic motion model. \label{tab:best_periodic} (table7.tex)

Next, we searched for evidence of periodicity using the Lomb-Scargle (LS) periodogram, which gives a least-square estimate of the periodogram based on unequally sampled time series data \citep{lomb1976,scargle1982}. We derived the LS periodograms for residuals of linear fittings in both RA and Dec directions. Uncertainties in the LS periodograms were estimated with the non-parametric percentile bootstrap method (e.g., \cite{akiyama2013}), which is a straightforward and efficient method in deriving estimates of confidence intervals particularly for large number of parameters. We created $10^5$ datasets (so-called bootstrap samples) by re-sampling, in which repetition of data was allowed. Each resulting dataset has the same number of data points as the original one. Periodigrams for all of $10^5$ bootstrap samples were calculated and then used to estimate confidence intervals. Percentile bootstrap confidence limits of the power specrum at each frequency were obtained as the edges of the middle 99.7\% (3$\sigma$) fractions of the bootstrap estimates.

\begin{figure}
\includegraphics[width=1.00\columnwidth]{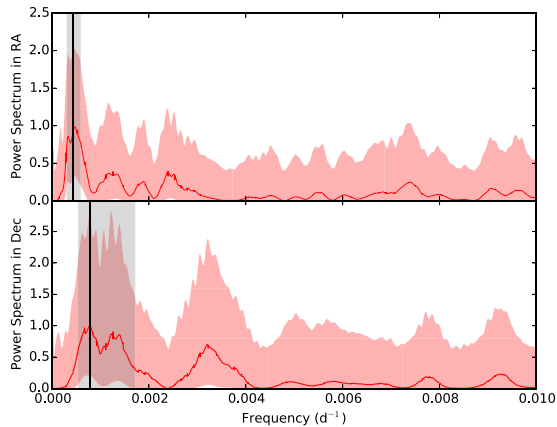}
\caption{
The Lomb-Scargle periodograms for residuals of linear-fittings in RA (top panel) and Dec (bottom panel) directions. Red solid lines indicate the least-square estimate of the power spectrum normalized by the maximum value, while their 3$\sigma$ uncertainties are shown in surrounding red-shaded regions, which are derived with the non-parametric percentile bootstrap method. Black vertical lines indicate the peaks of the 3$\sigma$-credible areas with the largest power and SNR, while gray-shaded regions exhibit their 3$\sigma$ uncertainties. 
\label{fig:LSperiod}}
\end{figure}

The derived LS periodograms for residuals of linear fittings are shown with their 3$\sigma$ uncertainties in figure \ref{fig:LSperiod}. A few regions are marginally detected with $>3\sigma$ in both RA and Dec directions, although most of powers pectrums are dominated by 3$\sigma$ errors. In RA directions, the peak with the largest power and signal-to-noise ratio (SNR) was located in a $3\sigma$-credible region at a frequency of $\sim 4.4\times10^{-4}$ d$^{-1}$ with significance of 4.3$\sigma$. The 3$\sigma$ estimate  of the peak frequency for this region is $4.4^{+1.5}_{-1.4}\times10^{-4}$ d$^{-1}$ corresponding to a period of $6.3^{+1.7}_{-2.8}$\ yr.
On the other hand, in Dec directions, the LS periodogram has the peak with the largest power and SNR in a $3\sigma$-credible region at a frequency of $\sim7.8\times10^{-4}$ with significance of 3.8$\sigma$. The 3$\sigma$ estimate of the peak frequency for this region is $7.8^{+94}_{-2.4}\times10^{-4}$ d$^{-1}$ corresponding to a period of $3.5^{+1.6}_{-1.9}$\ yr. 
The peak frequencies of periodograms are consistent between RA an Dec directions, and also with the derived period for the periodic motion model.

Physical parameters derived through the above two methods are given in table\ \ref{tab:fitting_function_and_model_selection}.
The period of the periodic motion in the source frame is calculated as $P_{\mathrm{source}}=\frac{P_{\mathrm{obs}}}{1+z}$, where $z$ is the source redshift, and $P_{\mathrm{obs}}$ is the period measured in the observer's frame.
The time span of our dataset ($\sim 6.2$\ yr) is comparable to these derived periods.
Therefore, we need additional monitoring of C3 motion to verify the periodic trend more precisely.

% table: Results of periodicity analyses. \label{tab:fitting_function_and_model_selection} (table8.tex)

\section{Discussion}
\subsection{The Nature of C3}
Observationally, the advance speed of hot spots of several CSOs are sub-relativistic ($\sim$ 0.1--0.3$c$), indicating dynamical ages of $\sim 10^{2}$--$10^{4}$\ yr taking account of their size of $\leq 1$\ kpc (e.g., \cite{Polatidis+1999}; \cite{Conway2002}; \cite{Nagai+2006}).
Theoretically, hot spot velocity ($v_{\mathrm{HS}}$) in the initial phase (\textit{one-dimensional dynamical evolution phase}; $t < 1.2 \times 10^{5}$\ yr) is nearly constant \citep[and references therein]{Kawakatu+Kino2006}.
Considering that the resultant velocity of C3 in 3C 84 is sub-relativistic ($\sim 0.40c$) and almost constant in a subparsec-scale jet in the initial phase ($\lesssim 10$\ yr), C3 shows similar behaviors to a terminal hot spot in a mini-radio lobe.
\footnote{More strictly speaking, the new radio component C3 is the head of radio lobe including hot spots at a very early stage of radio lobe evolution, since the higher resolution image using 43\ GHz VLBA revealed that the region around C3 showed very complex structure \citep{Nagai+2014}.
Similarity of the velocity of C4 ($\sim0.44c$) to that of C3 might mean that C4 is also the head of a mini-radio lobe including hot spots.
Although hot spots themselves cannot be resolved by VERA, our result implies that the radio lobes in radio galaxies might be already formed in subparsec-scale jets close to the central SMBHs.}

It is also worthwhile to emphasize that the measured advance speed of C3 assuming $\theta = \timeform{25D}$ is slightly faster than other CSOs.
This trend is also identified in the $\sim 15$\ mas ($\sim 5$\ pc) scale radio jet/lobe associated with the 1959 outburst in 3C 84 (\cite{Asada+2006}; \cite{Nagai+2008}).
The apparent speed of the hot spot on $\sim 5$-pc scale was $0.34\pm0.09c$ in 2001.
\citet{Nagai+2008} noticed that the hot spot on $\sim 5$-pc scale (component `B3' in \cite{Asada+2006}) was probably produced by the interaction between the jet ejected before 1959 and new-born jet components in the 1959 outburst, rather than by the interaction between the jet and ambient medium.
Similarly, the slightly faster speed of C3 on a subparsec scale ($\sim0.4c$) measured in the VERA monitoring might be the result of the interaction between the jet ejected before 2005 and newly ejected jet components in the 2005 outburst.

\subsection{Origin of the non-linear motion of C3}

Here we briefly discuss a possible origin of the C3 motion measured by VERA.
The possible periodic motion of C3 can be explained if the underlying continuous jet flow shows precession.
Precessions on subparsec-scale jets are generally caused by several physical mechanisms such as jet plasma instabilities, gravitational torques in a binary black hole system, magnetic torques, and accretion disk precession (e.g., \cite{Lobanov+Zensus2001}; \cite{Lobanov+Roland2005}; \cite{Mckinney+2013}; \cite{Caproni+2004}).

Since there is no evidence of a binary black hole system in the center, the most probable origin causing jet precession in 3C 84 is accretion disk precession by the Bardeen-Petterson  (BP) effect \citep{Bardeen+Petterson1975} acting on the viscous accretion disk originating the jet, which is tilted with regard to the equatorial plane of the central Kerr black hole, and inducing the alignment of the disk and the black hole angular momenta.
On 10--100\ kpc-scales in the X-ray band, the misaligned morphology is interpreted as a product of a precessing jet with a period of $3.3\times10^{7}$ years and semi-aperture angle of about \timeform{50D} \citep[and references therein]{Dunn+2006}. \citet{Falceta-Goncalves+2010} indeed showed that the observed morphology of 3C 84 on 10--100\ kpc-scales can be well explained by a precessing jet with a period of $5\times 10^{7}$\ years, using three-dimensional numerical simulations considering the jet precession evolution due to the BP effect.

Interestingly, Lister et al. (2013) also found similar 
significant changes in the innermost position angles of various 
blazars monitored in the MOJAVE project. 
They found that  there is some evidence of oscillatory behavior,
but the fitted periods (5-12 yr) are too long compared to
the length of the data set to firmly establish periodicity.
Although in the paper of Lister et al. (2013), the authors insist that
the measured periods are very short compared to expected precession
timescales from the BP effect.
However, the precession timescale due to the BP effect can be 
short enough at the late phase of 
the precession \citep{Scheuer+Feiler1996}.
Therefore, such precession phenomena may be ubiquitous in AGN jets
and could be understood as the BP-effect, although other possibilities cannot be ruled out.

It is worth to mention possible origins of this non-linear motion other than periodic motion.
Mizuta et al. (2010) pointed out that backflows generated at the jet head
can make influences on the jet itself.
When the head propagation velocity of the jet is smaller than the local sound speed, 
a bent backflow appears and it beats the jet from the sideways. 
Such influences of backflows can potentially explain the detected
non-linear motion in the 3C 84 jet.
Non-uniform density distributions of the surrounding cocoon seen in various
hydrodynamical simulations of relativistic jet propagations (e.g., Scheck et al. 2002)
may also contribute to non-linear motion.

\section{Summary}
\citet{Suzuki+2012} found that the subparsec-scale jet component C3 had emerged from the radio core before 2005, and traveled southward following a parabolic trajectory on the celestial sphere with VLBA at 43\ GHz from 2003 November to 2008 November.
In this paper, we further explored the kinematics of C3 from 2007 October to 2013 December (80 epochs) using 22\ GHz VERA data.
Summary and discussions are as follows.

\begin{itemize}
\item We find that the averaged apparent speed of C3 relative to the radio core is almost constant and sub-relativistic ($0.27\pm0.02c$) from 2007 October to 2013 December.
This property suggests that C3 may be the head of a mini-radio lobe including hot spots, rather than a bright knotty component in an underlying continuous jet flow.
This result implies that the radio lobe in radio-loud AGNs might be already formed in subparsec-scale jets in the vicinity of SMBHs.

\item Although the observation time span was not enough to derive a final conclusion,
we find a possible helical path  of C3 with a period of about six years. 
Although we cannot reliably identify the origin of the wobbling motion due to the insufficient time span of our dataset and the lack of information about the absolute reference position, the motion might reflect a precessing jet nozzle, induced by the Bardeen-Petterson effect.
In order to obtain more robust results, we continue to monitor the subparsec-scale jet of 3C 84 with high resolution (phase-referencing and polarization) VLBI.
\footnote{To avoid a possible confusion for readers, we should note that recent VLBI 
observations show the flip of  C3 position in 2016 autumn (Nagai et al. 2017; Kino et al. 2018),
although the flip was an instantaneous phenomenon and it is independent from the result reported in this work.}

\end{itemize}
\bigskip

%%%%%%%%%%%%%%%%%%%%%%%%%%%%%%%%%%%%%%%

\begin{ack}
%We appreciate the detailed review and useful suggestions of anonymous referee who have improved the original manuscript.
We are grateful to all staff of the VERA stations for their assistance in observations. 
We thank the anonymous referees for their helpful comments. 
K.H. would like to thank M. Umei and Y. Nishikawa for helpful comments on statistical analyses.
K.H. also thanks Y. Fujimoto and all members of the Astrophysical laboratory, Hokkaido University for encouragement to carry out this work.
VERA is operated by the National Astronomical Observatory of Japan.
\end{ack}

%\appendix 
%\section*{Case of single paragraph}

%\section{Case of two or paragraphs}

%\section{Case of two or paragraphs}

\newpage

% supplementary tables
%\input{supplementary_table1}
%\input{supplementary_table2}
%\input{supplementary_table3}
%\input{supplementary_table4}
%\input{supplementary_table5}

%tables in the main manuscript
\begin{table}
\caption{The best-fit parameters of the linear motion model.}
\label{tab:best_linear}
\begin{tabular}{lc}
\hline\hline
Paramameter & Best-fit value \\
\hline
$a (\mathrm{mas\ day^{-1}})$ & $(-1.4 \pm 2.6)\times 10^{-5}$ \\
$b (\mathrm{mas})$ & $0.10 \pm 0.04$ \\
$c (\mathrm{mas\ day^{-1}})$ & $(-6.4 \pm 0.5)\times 10^{-4}$ \\
$d (\mathrm{mas})$ & $-0.80 \pm 0.08$ \\
\hline
\multicolumn{2}{@{}l@{}}{\hbox to 0pt{\parbox{85mm}{\footnotesize
Notes.
\par\noindent
Errors are $3\sigma$-confidence intervals estimated by the Monte Carlo method with $10^{5}$ trials.
}\hss}}
\end{tabular}
\end{table}
\begin{table}
\caption{The best-fit parameters of the periodic motion model.}
\label{tab:best_periodic}
\begin{tabular}{lc}
\hline\hline
Paramameter & Best-fit value \\
\hline
$A (\mathrm{mas})$ & $0.050_{-0.023}^{+0.20}$ \\
$P_{\mathrm{obs}}^{\mathrm{fit}} (\mathrm{days})$ & $(2.0_{-0.4}^{+3.1})\times 10^{3}$ \\
$B (\mathrm{deg})$ & $-95_{-85}^{+160}$ \\
$C (\mathrm{mas\ day^{-1}})$ & $(3.8_{-28}^{+210})\times 10^{-6}$ \\
$D (\mathrm{mas})$ & $0.09_{-0.31}^{+0.04}$ \\
$E (\mathrm{mas})$ & $0.07_{-0.05}^{+0.51}$ \\
$F (\mathrm{deg})$ & $-45_{-90}^{+140}$ \\
$G (\mathrm{mas\ day^{-1}})$ & $(-6.1_{-0.7}^{+5.5})\times 10^{-4}$ \\
$H (\mathrm{mas})$ & $-0.80_{-0.66}^{+0.10}$ \\
\hline
\multicolumn{2}{@{}l@{}}{\hbox to 0pt{\parbox{85mm}{\footnotesize
Notes.
\par\noindent
Errors are $3\sigma$-confidence intervals estimated by the Monte Carlo method with $10^{5}$ trials.
}\hss}}
\end{tabular}
\end{table}
\begin{table*}
\caption{Results of periodicity analyses.}
\label{tab:fitting_function_and_model_selection}
\begin{tabular}{lcccc}
\hline\hline
Method & \multicolumn{2}{c}{Period (yr)\footnotemark[$*$]} & \multicolumn{2}{c}{Amplitude ($\times 10^{-2}\ \mathrm{pc}$)\footnotemark[$\dagger$]} \\
 & R.A. & Decl. & R.A. & Decl. \\
\hline
Best fit & \multicolumn{2}{c}{$5.3_{-1.2}^{+8.3}$} & \multicolumn{1}{c}{$1.8_{-0.8}^{+6.9}$} & \multicolumn{1}{c}{$2.5_{-1.9}^{+17.9}$} \\
Lomb-Scargle & $6.2_{-2.8}^{+1.7}$ & $3.4_{-1.9}^{+1.6}$ & \multicolumn{1}{c}{$\cdots$} & \multicolumn{1}{c}{$\cdots$} \\
\hline
\multicolumn{5}{@{}l@{}}{\hbox to 0pt{\parbox{90mm}{\footnotesize
Notes.
\par\noindent
Error estimation is described in the text in detail.
\par\noindent
\footnotemark[$*$] Period of the periodic motion measured in the source frame, $P_{\mathrm{source}}=\frac{P_{\mathrm{obs}}}{1+z}$, where $z$ is the source redshift, and $P_{\mathrm{obs}}$ is the period measured in the observer's frame.
The period derived by the best fit method (least square method) is the common parameter between right ascension and declination directions.
\par\noindent
\footnotemark[$\dagger$] Amplitude of the periodic motion along each axis.
The method using Lomb-Scargle periodograms does not tell the information about amplitude.
}\hss}}
\end{tabular}
\end{table*}

%%%%%%%%%%%%%%%%%%%%%%%%%%%%%%%%%%%%%%%%%%%%%%%%%%%%%%%%%%%%%%%%%%%%%%
%%%
% See the manual for the detail.
%%%

\end{document}